\documentclass[12pt, a4paper]{elsarticle}

\usepackage{amssymb}
\usepackage{hyperref}
\usepackage{sessiontypes}
\usepackage{xspace}
\usepackage{listings}
\usepackage{fancyvrb,fvextra}
\usepackage{cprotect}
\usepackage{tikz}
\usepackage{mathtools}
\usepackage{bm}
\usepackage{amsfonts}
\usepackage{amsmath}
\usepackage{amsthm}
\usepackage{cleveref}
\usepackage{wrapfig}
\usepackage{subcaption}
\usepackage{extpfeil}
\usepackage{etoolbox}
\usepackage{xcolor}
\usepackage{booktabs}
\usepackage{cases}
\usepackage{semantic}
\usepackage{relsize}
\usepackage{pgf}
\usepackage{xstring}
\usepackage{xspace}
\usepackage{pifont}
\usepackage{nicefrac}
\usepackage[symbol]{footmisc}

\usepackage[inline,shortlabels]{enumitem}

\makeatletter
\@ifclasswith{elsarticle}{preprint}{%
  \makeatletter
  \patchcmd{\@addmarginpar}{\ifodd\c@page}{\ifodd\c@page\@tempcnta\m@ne}{}{}
  \makeatother
}{%
}
\makeatother

\newcommand{\cmark}{\ding{51}}

\definecolor{yellowhighlight}{RGB}{255, 255, 64}
\definecolor{orangehighlight}{RGB}{255, 204, 153}

\newcommand*{\eg}{\textit{e.g.,}\xspace}

\newcommand*{\ie}{\textit{i.e.,}\xspace}
\newcommand*{\etc}{\textit{etc.}\xspace}

\newcommand*{\wrt}{w.r.t.\xspace}

\newcommand{\lchannels}{\texttt{lchannels}\xspace}

\newcommand{\stmonitor}{\texttt{STMonitor}\xspace}

\definecolor{dkgreen}{rgb}{0,0.6,0}
\definecolor{gray}{rgb}{0.5,0.5,0.5}
\definecolor{mauve}{rgb}{0.58,0,0.82}

\usetikzlibrary {
  matrix,
  shapes,
  arrows,
  shadows,
  calc,
  chains,
  decorations.pathmorphing,
  decorations.text,
  arrows.meta,
  patterns,
  fit,
  bending,
  shadows,
  shapes.geometric,
  snakes
}

\tikzset{
  label/.style={
    font=\scriptsize\itshape,
    inner sep=0.4em
  },
  point/.style={
    circle,
    fill=black,
    text width=0.3em,
    inner sep=0
  },
  state/.style={
    circle,
    text width=0.8em,
    inner sep=0.1em,
    text depth=0.08em,
    draw,
    font=\scriptsize
  },
  participant/.style={
    draw=black,
    rounded corners,
    semithick,
    font=\footnotesize,
    text height=0.2cm,
    text centered,
    anchor=base
  },
  specification/.style={
    font=\small,
    semithick,
    draw=black
  },
  synthesiser/.style={
    draw=black,
    dotted,
    semithick,
    inner sep=2pt,
    font=\footnotesize
  },
  monitor/.style={
    draw=black,
    fill=white,
    semithick,
    minimum width=0.5cm,
    minimum height=0.5cm,
    font=\footnotesize,
    text height=0.2cm,
    drop shadow=ashadow
  },
  point/.style={
    minimum size=0pt, 
    inner sep=0pt
  },
  ltspoint/.style={
    circle,
    fill=black,
    draw=white,
    line width=0.5mm,
    text width=0.25em,
    inner sep=0,
    font=\footnotesize
  },
  ashadow/.style={
    opacity=.3, 
    shadow xshift=0.6mm, 
    shadow yshift=-0.6mm 
  },
  oppashadow/.style={
    opacity=.3, 
    shadow xshift=-0.6mm, 
    shadow yshift=-0.6mm 
  },
  process/.style={
    draw=black,
    fill=white,
    semithick,
    minimum width=0.5cm,
    minimum height=0.5cm,
    font=\footnotesize,
    text height=0.2cm,
    drop shadow=ashadow
  },
  event/.style={
    draw=black,
    semithick,
    minimum width=0.4cm,
    minimum height=0.4cm,
    font=\scriptsize
  },
  trap/.style={
    trapezium, 
    trapezium angle=67.5, 
    draw,
    inner ysep=5pt, 
    outer sep=0pt,
    minimum height=1.81mm, 
    minimum width=0pt
  },
  static-process/.style={
    draw=black,
    fill=white,
    semithick,
    minimum width=0.5cm,
    minimum height=0.5cm,
    font=\footnotesize,
    text height=0.2cm
  },
  pill/.style={
    draw,
    text=black,
    fill=white,
    minimum width=0.84em, 
    minimum height=0.8em,
    inner sep=0.1em,
    outer sep=0,
    font=\scriptsize\sffamily,
    align=center,
    rounded corners=0.4em  
  }
}

\pgfdeclaredecoration{penciline}{initial}{
    \state{initial}[width=+\pgfdecoratedinputsegmentremainingdistance,auto corner on length=1mm,]{
        \pgfpathcurveto%
        {
            \pgfqpoint{\pgfdecoratedinputsegmentremainingdistance}
                            {\pgfdecorationsegmentamplitude}
        }
        {
        \pgfmathrand
        \pgfpointadd{\pgfqpoint{\pgfdecoratedinputsegmentremainingdistance}{0pt}}
                        {\pgfqpoint{-\pgfdecorationsegmentaspect\pgfdecoratedinputsegmentremainingdistance}%
                                        {\pgfmathresult\pgfdecorationsegmentamplitude}
                        }
        }
        {
        \pgfpointadd{\pgfpointdecoratedinputsegmentlast}{\pgfpoint{1pt}{1pt}}
        }
    }
    \state{final}{}
}

\newcounter{example}[section]
\newenvironment{example}[1][]{\refstepcounter{example}\par\smallskip
   \noindent \textbf{Example~\theexample. #1}\rmfamily}{\smallskip}

\definecolor{dkgreen}{rgb}{0,0.6,0}
\definecolor{gray}{rgb}{0.5,0.5,0.5}
\definecolor{mauve}{rgb}{0.58,0,0.82}

\makeatletter
\def\ps@pprintTitle{%
 \let\@oddhead\@empty
 \let\@evenhead\@empty
 \def\@oddfoot{}%
 \let\@evenfoot\@oddfoot}
\makeatother

\usepackage[index,nomargin,inline,%
  status=draft
]{fixme} 
\fxusetheme{colorsig}%
\FXRegisterAuthor{fxAF}{anfxAF}{}
\FXRegisterAuthor{fxAS}{anfxAS}{}
\FXRegisterAuthor{fxCBB}{anfxCBB}{}
\FXRegisterAuthor{fxCT}{anfxCT}{}
\FXRegisterAuthor{fxeM}{anfxeM}{}

\newcommand{\cpspc}{\textsf{CPSPc}\xspace}

\newcommandx{\cm}[1][1={}]{$\textsf{CM}_{#1}$\xspace}
\newcommand{\cmclient}{$\textsf{CM}_{c}$}

\newcommand{\cms}{$\textsf{CM}$s\xspace}

\newcommand{\pstmonitor}{\texttt{PSTMonitor}\xspace}
\newcommand{\sgame}{$\stS[game]$\xspace}
\newcommand{\server}{\emph{s}\xspace}

\journal{Software Impacts}

\begin{document}

\lstset{frame=tb,
  language=Scala,
  showstringspaces=false,
  columns=flexible,
  basicstyle={\linespread{1}\footnotesize\ttfamily},
  numbers=left,
  numberstyle=\tiny\color{gray}\sffamily,
  keywordstyle=\color{blue},
  commentstyle=\color{dkgreen},
  stringstyle=\color{mauve},
  breaklines=true,
  breakatwhitespace=true,
  tabsize=3,
}

\begin{frontmatter}

\title{\pstmonitor: Monitor Synthesis from\\Probabilistic Session Types\footnote[1]{The final authenticated version is available online at: \url{https://doi.org/10.1016/j.scico.2022.102847}. The work has been partly supported by: the project MoVeMnt (No:\,217987-051) under the Icelandic Research Fund; the BehAPI project funded by the EU H2020 RISE under the Marie Sk{\l}odowska-Curie action (No:\,778233); the MIUR projects PRIN 2017FTXR7S IT MATTERS and 2017TWRCNB SEDUCE; the EU Horizon 2020 project 830929 \textit{CyberSec4Europe}; the Danish Industriens Fonds Cyberprogram 2020-0489 \textit{Security-by-Design in Digital Denmark}.}}

\author{Christian Bartolo Burl\`{o}\textsuperscript{1}}
\author{Adrian Francalanza\textsuperscript{2}}
\author{Alceste Scalas\textsuperscript{3}}
\author{\\Catia Trubiani\textsuperscript{1}}
\author{Emilio Tuosto\textsuperscript{1}}
\address{\textsuperscript{1}Gran Sasso Science Institute, Italy, \textsuperscript{2}Department of Computer Science, University of Malta, Malta, \textsuperscript{3}DTU Compute, Technical University of Denmark, Denmark\vspace{-1cm}}

\begin{abstract}
  We present \pstmonitor, a tool for the run-time verification of quantitative specifications of
  message-passing applications, based on probabilistic session types.
  The key element of \pstmonitor is the detection of executions
  that deviate from expected probabilistic behaviour.
  Besides presenting \pstmonitor and its operation, the paper analyses its feasibility in terms of the runtime overheads it induces.
\end{abstract}


\end{frontmatter}


\section{Introduction}

This paper showcases \pstmonitor, a tool supporting the run-time monitoring of quantitative properties of
message-passing applications.
The underlying methodology of \pstmonitor, recently described in the
companion paper~\cite{DBLP:conf/coordination/BurloFSTT21}, relies on
\emph{(probabilistic) session types} (PST for short).
%
PSTs  specify both \begin{enumerate*}[label=\textit{(\roman*)}] \item the
expected communication protocol; and \item quantitative constraints on the
branching behaviour of the protocol through probabilities, within one unified formalism. \end{enumerate*}
%

\begin{example}\label{eg:example-intro}
  Consider a server hosting a guessing game. The server picks an integer $n$ between 1 and 100,
  and the client is expected to either attempt to guess $n$, or ask for a hint.
  In such setup, one might want to assess whether the server behaves fairly and gives the client a chance of guessing correctly. 
  On the other end, one might also want to assess whether the client is asking for too many hints without attempting to guess.
  \qed
\end{example}

The methodology proposed in~\cite{DBLP:conf/coordination/BurloFSTT21} explores the post-deployment verification of systems such as the one in \Cref{eg:example-intro} against PSTs.
In particular, monitors are used to detect deviations of the current run-time execution of a system from the probabilistic behaviour specified by the PST.
Given a PST, our tool \pstmonitor 
automates the synthesis of a
monitor that is able to both \begin{enumerate*}[label=\textit{(\roman*)}] \item report violations in case of miscommunications; and
\item issue warnings if the observed communications deviate substantially from the probabilistic 
behaviour expressed in the PST.
\end{enumerate*}
One complication is that PSTs specify probabilistic behaviours over \emph{complete} executions,
whereas our synthesised monitors need to detect deviations in real-time,
hence they can only make decisions based on a \emph{partial} execution observed up to the current instant.
For this reason, \pstmonitor\ relies on \emph{confidence intervals} to 
gauge whether the observed behaviour (up to the current execution point) is compatible with a given PST.

The rest of the paper is structured as follows.
\Cref{sec:workflow} illustrates key functionalities of our tool and
the workflow for deploying monitors derived from PSTs.
\Cref{s:evaluation} discusses the feasibility of our probabilistic
monitoring by considering a fragment of the Simple Mail Transfer
Protocol (SMTP)~\cite{SMTP}.
Finally, \Cref{sec:conc} draws some conclusions and sketches future
work.  

\section{Workflow and Tool Functionality}\label{sec:workflow}

The workflow of \pstmonitor starts with a client-server communication protocol expressed as a \emph{probabilistic session type (PST)} \cite{DBLP:journals/corr/abs-2011-09037, DBLP:conf/concur/InversoMPTT20} --- \ie a (binary) session type where the choice-points are augmented with quantitative information, namely probabilities describing the frequencies of the choices to be taken.
The PST is then used to automatically synthesise a monitor 
which operates \wrt a given \emph{confidence level}, as detailed in \cite[Section 2.2]{DBLP:conf/coordination/BurloFSTT21}. 
Intuitively, the synthesised monitor iteratively calculates \emph{confidence intervals} around all choice-points specified in the PST, and computes frequency-based estimates of the probabilities of the choices observed in the monitored system. The monitor continuously checks whether the run-time-observed choice probabilities fall within the corresponding PST confidence intervals: if not, the monitor issues a warning, meaning that
the observed behaviour is significantly deviating from the PST specification.
The monitor can later retract the warning, if the run-time-observed choice probabilities return within the PST confidence intervals.

\subsection{Specifying behaviour via Probabilistic Session Types (PSTs)}

Session types formalise communication protocols by specifying the order and choice of messages together with their corresponding payloads. 
We take \emph{binary} session types (supporting client-server protocols) and augment the choice points with probability distributions that specify the frequency with which a choice should be taken by one of the components interacting in a session. 
The syntax of our Probabilistic Session Types (PSTs) is thus:
\begin{align*}
  \stS \;\bnfdef\; &\ \stBrai{i}{I} && \textit{(external choice)} \\
    \bnfmid & \stSeli{i}{I} && \textit{(internal choice)} \\
    \bnfmid & \stRec{X}.\stS && \textit{(recursion)} \\ 
    \bnfmid & \stRecVar{X} && \textit{(recursion variable)} \\
    \bnfmid & \stEnd && \textit{(termination)}
\end{align*}
In choice points ($\stBraOp$ and $\stSelOp$) the indexing set $I$ is finite and non-empty, the \emph{choice labels} $\texttt{l}_i$ are pairwise distinct, and the \emph{sorts} $\asort_i$ range over basic data types ($\typeInt$, $\typeStr$, $\typeBool$,\etc) for typing \emph{variables} $x_i$. 
We give a \emph{multinomial distribution} interpretation to each choice point ($\stBraOp$ and $\stSelOp$) in a PST:
we require that $\sum_{i \in I} p_i = 1$, where every $0 \leq p_i \leq 1$ is the probability of selecting the branch labelled by $\texttt{l}_i$.
The probabilities prescribed at a choice point represent a behavioural obligation on the interacting party that has control over the selection at that choice point. 
As usual, we require that recursion is guarded, \ie a recursion variable $X$ can only appear under an external or internal choice prefix.

\begin{example}
  \label{eg:session-type}
  We formalise the communication protocol outlined in \Cref{eg:example-intro} as the PST $\stS[game]$ in \Cref{fig:s-game}, written from the perspective of the server. 
  The type specifies that the server should wait for the client's choice (at the external branching point \stBraOp) to either \lab{Guess} a number, ask for \lab{Help}, or \lab{Quit}. 
  If the client asks for help, the server should reply with a \lab{Hint} message including a string, and the session loops. 
  The session should also loop after the outcome of a guess (\lab{Correct} or \lab{Incorrect}, at the internal choice \stSelOp) is communicated to the client. 
  
  The probability annotations in $\stS[game]$ specify the expected frequency of each choice, and rule out unwanted behaviours. 
  In particular, they require the server to reply with \lab{Correct} 1\% of the time, and the client to only request for help 20\% of the time.
  \qed
\end{example}

\begin{figure}[h]
  \small\begin{Verbatim}[commandchars=\\\{\},numbers=left,numbersep=1mm,frame=single,breaklines]
    S_game = rec X.(+\{!Guess(num: Int)[0.75].
                    &\{?Correct()[0.01].X, ?Incorrect()[0.99].X\},
                  !Help()[0.2].?Hint(info: String)[1].X,
                  !Quit()[0.05].end\})
  \end{Verbatim}
  \vspace{-0.5cm}\cprotect\caption{Probabilistic session type \verb|S_game|.}\label{fig:s-game}
\end{figure}

\subsection{Monitor synthesis: behind the scenes}

\begin{figure}[htp]
  \centering
  \begin{subfigure}{0.45\textwidth}
    \centering
    \begin{tikzpicture}[decoration=penciline]
      \node(monitor) [static-process]{\texttt{Monitor}};
      
      \node(synth) [static-process, above=0.5cm of monitor, font=\footnotesize] {$\textsf{synth}$};
      \node(cpspc) [draw, static-process, right=0.5cm of synth, font=\footnotesize] {$\textsf{CPSPc}$};
      \node(scb) [decorate, draw, left=0.7cm of synth] {$\stS[game]$};

      \draw[-angle 90,dashed] (scb) edge (synth);
      \draw[-stealth, semithick] (synth) edge (monitor);
      \draw[-stealth, semithick] (synth) edge (cpspc);
    \end{tikzpicture}
    \caption{\centering Monitor and CPSPc synthesis.}\label{fig:mon-cpspc-synth}
  \end{subfigure}
  \hspace{0.03\textwidth}
  \hspace{0.03\textwidth}
  \begin{subfigure}{0.45\textwidth}
    \centering
    \begin{tikzpicture}[decoration=penciline]
      \node(monitor) [static-process]{\texttt{Monitor}};
      \node(cm_c) [draw, left=0.2cm of monitor, font=\scriptsize] {$\textsf{CM}_c$};
      \node(cm_m) [draw, right=0.2cm of monitor, font=\scriptsize] {$\textsf{CM}_s$};
      \node(cpspc) [draw, static-process, above=0.5cm of monitor, font=\footnotesize] {$\textsf{CPSPc}$};
      
      \node(scala-typechecker)[draw, right=1cm of cpspc, inner sep=2pt, font=\scriptsize, align=center]{\texttt{Scala}\\[-1mm]\texttt{type-checker}\\[-1mm]\texttt{+ lchannels}};

      \draw[-angle 90, dashed] (cpspc) --  (scala-typechecker);
      \draw[-, dashed] (cm_c) |- ($(cm_m.north)+(0,0.35)$);
      \draw[-, dashed] (cm_m) |- (scala-typechecker);
      \draw[-, dashed] (monitor) -- ($(monitor.north)+(0,0.35)$);

      \node(satisfaction) [circle, draw, below=0.4cm of scala-typechecker, text=green!70!black, font=\scriptsize, inner sep=1pt, minimum size=4mm, semithick] {\cmark};
      \draw[-stealth, semithick] (scala-typechecker) edge (satisfaction);
    \end{tikzpicture}
    \caption{\centering Type-checking implementations.}\label{fig:type-checking-implementations}
  \end{subfigure}
  \caption{Synthesising and compiling the monitor.}\label{fig:synth-compile}
\end{figure}
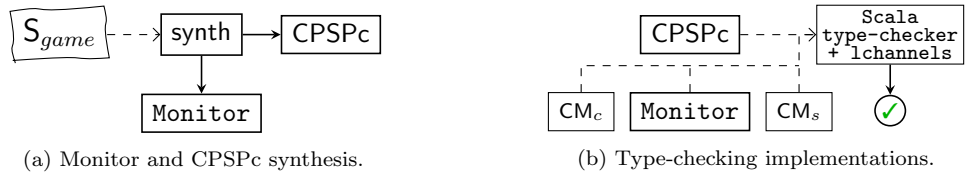

\pstmonitor generates executable session monitors in the Scala programming language.
However, session type constructs (internal or external choices) are not natively supported by Scala (nor other mainstream programming languages). 
For this reason, \pstmonitor leverages the library \lchannels \cite{DBLP:conf/ecoop/ScalasY16}, which encodes (binary) session types constructs in the Scala type system through \emph{Continuation-Passing Style Protocol classes} (\cpspc), capturing the order of the send and receive operations and choices in a session type.

\begin{example}
As depicted in \Cref{fig:mon-cpspc-synth}, our tool \pstmonitor generates the \cpspc from a session type.
\Cref{lst:cpspc} contains some of the \cpspc from the autogenerated file representing the type \sgame from \Cref{eg:session-type}. 
\begin{lstlisting}[caption={Continuation-Passing Style Protocol classes for \sgame},label={lst:cpspc},captionpos=b,aboveskip=3mm]
sealed abstract class ExtChoice1
case class Guess(num: Int)(val cont: Out[IntChoice1]) extends ExtChoice1
sealed abstract class IntChoice1
case class Correct()(val cont: Out[ExtChoice1]) extends IntChoice1 
case class Incorrect()(val cont: Out[ExtChoice1]) extends IntChoice1
// ... remaining classes representing the session type 
\end{lstlisting}
Intuitively, every class represents a message specified within the session type (lines 2, 4 and 5), containing the 
\begin{enumerate*}[label=\emph{(\roman*)}]
  \item type of the payload; and 
  \item the continuation type. 
\end{enumerate*}
Every choice point is represented by an abstract class (lines 1 and 3) that is inherited by all the choices within the choice point. 
The monitor uses these types to keep track of the current point of the interaction and verify that the components respect the session type. \qed
\end{example}

\begin{wrapfigure}[11]{l}{8.9cm}
  \vspace{-0.6cm}
\begin{lstlisting}[caption={\vspace{1cm}\texttt{receive} and \texttt{send} methods of \cmclient~in \Cref{fig:starting-monitor}.},label={lst:cm-receive},captionpos=b,aboveskip=3mm,frame=tbr,rulecolor=\color{black},firstnumber=1]
val guessR = """GUESS (.*)""".r
def receive(): Any = inBuf.readLine() match {
  case guessR(num) => Guess(num.toInt)(null)
  // case ...
  case other => other
}
def send(msg: Any): Unit = msg match {
  case Correct() => outB.write(f"CORRECT\n")
  case Incorrect() => outB.write(f"INCORRECT\n")
  // case ...
  case _ => { close()
               throw Exception("Invalid msg") }
}
\end{lstlisting}
\end{wrapfigure} 
\noindent The monitors generated by \pstmonitor are agnostic to the transport protocol in use:
they require user-supplied Connection Managers (\cms) that sit between them and the components under observation (as depicted in \Cref{fig:type-checking-implementations}). 
A \cm acts as a translator and gatekeeper by transforming messages from the format used in the transport protocol to their respective \cpspc (and \emph{vice versa}), while governing the interaction with the component. 
In order to do this, \cms must extend a provided abstract class (\verb|ConnectionManager|) and implement the methods \verb|setup|, \verb|close|, \verb|receive| and \verb|send| (as in \Cref{lst:cm-receive}). 
As the name implies, the first two manage the connection with the component; the monitor invokes them when it sets up the connection and before it terminates.

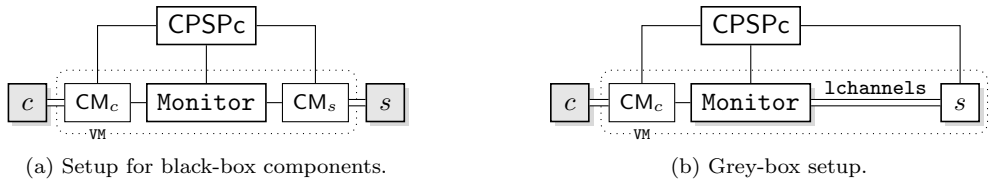
\begin{figure}[htp]
  \centering
  \begin{subfigure}{0.45\textwidth}
    \centering
    \begin{tikzpicture}
      \node(monitor) [static-process]{\texttt{Monitor}};
      \node(cm_c) [draw, left=0.2cm of monitor, font=\scriptsize] {$\textsf{CM}_c$};
      \node(cm_m) [draw, right=0.2cm of monitor, font=\scriptsize] {$\textsf{CM}_s$};
      \draw [thin, dotted, rounded corners] ($(cm_c.north west)+(-0.1cm,0.17cm)$) rectangle ($(cm_m.south east)+(0.1cm,-0.15cm)$);
      \node(server) [process, fill=gray!20, right=1.3cm of monitor] {$s$};
      \node(client) [process, fill=gray!20, left=1.3cm of monitor] {$c$};

      \node(cpspc) [draw, above=0.5cm of monitor, static-process, font=\footnotesize] {$\textsf{CPSPc}$};

      \draw[-] (cm_c) -- (monitor);
      \draw[-] (cm_m) -- (monitor);
      \draw[-] (cpspc) -- (monitor);
      \draw[-] (cpspc) -| (cm_c);
      \draw[-] (cpspc) -| (cm_m);
      \draw[thin,double distance=2pt, fill=white] (cm_c) -- (client);
      \draw[thin,double distance=2pt, fill=white] (cm_m) -- (server);
      \node(vm) [below=0.8mm of cm_c, font=\tiny, draw=white, fill=white, inner sep=0pt] {\texttt{VM}};
    \end{tikzpicture}
    \caption{\centering Setup for black-box components.}\label{fig:wrapper-setup}
  \end{subfigure}
  \hspace{0.03\textwidth}
  \hspace{0.03\textwidth}
  \begin{subfigure}{0.45\textwidth}
    \centering
    \begin{tikzpicture}
      \node(client) [process, fill=gray!20] {$c$};
      \node(cm) [draw, right=0.25cm of client, font=\scriptsize] {$\textsf{CM}_c$};
      \node(monitor) [process, right=0.2cm of cm]{\texttt{Monitor}};
      \node(server) [process, right=1.7cm of monitor] {$s$};
      \draw [thin, dotted, rounded corners] ($(cm.north west)+(-0.1cm,0.17cm)$) rectangle ($(server.south east)+(0.1cm,-0.15cm)$);
      
      \draw[thin,double distance=2pt, fill=white] (cm) -- (client);
      \draw[thin,double distance=2pt, fill=white] (monitor) -- node(lchannels)[above, font=\scriptsize, yshift=-1pt]{\lchannels} (server);
      
      \node(cpspc) [draw, above=0.5cm of monitor, static-process, font=\footnotesize] {$\textsf{CPSPc}$};

      \draw[-] (cm) -- (monitor);

      \draw[-] (cpspc) --  (monitor);
      \draw[-] (cpspc) -|  (cm);
      \draw[-] (cpspc) -|  (server);
      \node(vm) [below=0.8mm of cm, font=\tiny, draw=white, fill=white, inner sep=0pt] {\texttt{VM}};
    \end{tikzpicture}
    \caption{\centering Grey-box setup.}\label{fig:hybrid-setup}
  \end{subfigure}
  \caption{Alternative monitoring setups for a client $c$ and a server $s$.}\label{fig:starting-monitor}
\end{figure}

\begin{example}
Listing \ref{lst:cm-receive} shows a code snippet for a \cm's \verb|receive| and \verb|send| methods.
In this case, the \cm is handling a TCP/IP socket, where the messages from the type $\stS[game]$ in \Cref{eg:example-intro} are serialised into a textual format, \eg $\lab{Guess}(n)$ into \texttt{"GUESS $n$"}.

\noindent When invoked by the monitor, the \verb|receive| method checks the socket input buffer \texttt{inBuf}: if it finds a supported message (line 3, where the message matches the regex \texttt{guessR}), it returns an object of the corresponding CPSP class; 
otherwise, it returns the unaltered message.
Therefore, when the client $c$ in \Cref{fig:starting-monitor} sends the message \texttt{"GUESS 23"}, the \cm[c] translates it to the CPSP class \texttt{Guess} (line 2 of \Cref{lst:cpspc}). 

When the server sends a message to the monitor to be forwarded to the client, it invokes the method \verb|send| in Listing \ref{lst:cm-receive}: such a method translates messages from a CPSP class instance into the format accepted by the client, and sends them. 
In this case, if the server replies with a message \texttt{Correct} (line 4 of Listing \ref{lst:cpspc}), the server translates it into the textual format \texttt{"CORRECT"} and writes it on the TCP/IP socket output buffer (line 7 of Listing \ref{lst:cm-receive}).
The catch-all case (line 10) is used for debugging purposes.
\qed 
\end{example}

\subsection{Synthesising and using a monitor}
\label{sec:synthesise-monitor}

Our tool \pstmonitor automatically generates the monitor code and the \cpspc from a session type description, by running the following command:
\begin{lstlisting}[language=bash]
$ sbt "project monitor" "runMain monitor.Generate $DIR $ST $PRE"
\end{lstlisting}
\begin{enumerate}[nosep]
  \item[\texttt{\$DIR}] is the directory where the source code of the monitor and classes will be generated;
  \item[\texttt{\$ST}] is the file containing the probabilistic session type (as in \Cref{fig:s-game}); and
  \item[\texttt{\$PRE}] (optional) is a file containing a preamble that will be added to the top of the generated files (\eg containing package declarations and imports).
\end{enumerate}
Once completed, the auto-generated files \texttt{Monitor.scala} and \texttt{CPSPc.scala} will be saved in the directory \texttt{\$DIR}. 

\subsection{Deploying the monitor}

The monitor generated in \Cref{sec:synthesise-monitor} can be used in a number of different setups. 
We outline two such setups covering two common situations, depicted in \Cref{fig:starting-monitor}:

\begin{description}[nosep]
  \item[Black-box monitoring setup (\Cref{fig:wrapper-setup}).]  This is the most general way to deploy a monitor: both the client $c$ and server $s$ are treated as black boxes. 
  Here the monitor makes use of two connection managers, one for each end of the interaction. 
  To start the monitor itself, the user needs to write a simple proxy that sits between the monitored client and server, and starts the generated monitor whenever a new connection is established. 
  \item[Grey-box monitoring setup (\Cref{fig:hybrid-setup}).]  This setup is possible when one of the components (\eg the server \server) is implemented in Scala using \lchannels and can be deployed together with the monitor.
  This allows for a direct \lchannels-based connection between the monitor and the component, without a connection manager. 
  Moreover, the monitor and component can be started on the same Java Virtual Machine (JVM),
  thus improving the overall performance (as we illustrate in \Cref{s:evaluation}).\footnote{%
    If a black-box component is supplied as a \texttt{.jar} file, it is also possible to deploy such a component and its monitor as different threads running on the same JVM; in this case, the monitor-component interaction would use a TCP/IP socket, which is less efficient than a direct \lchannels-based connection. The resulting performance would be similar to the black-box scenario measured in \Cref{fig:wrapper-setup}.}
\end{description}

\subsection{Additional Features}\label{s:monitoring-procedure}

To further assist with the analysis of a monitored system, monitors generated by \pstmonitor can also log information about the current execution, thus enabling, \eg the real-time visualisation of the monitor status.
The logs include \begin{enumerate*}[label=\emph{(\roman*)}]
  \item the estimated probability of how often a choice within a branch was taken; and
  \item the boundary of the confidence interval. 
\end{enumerate*} 

\begin{figure}
  \hspace{2.5em}
  \includegraphics[scale=0.11]{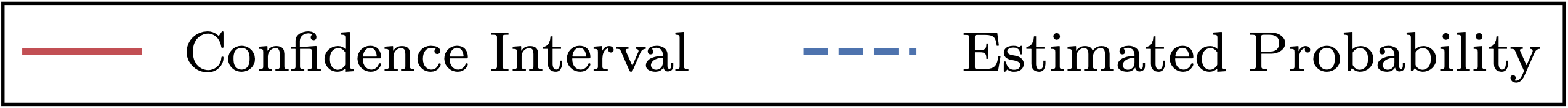}

  \smallskip
  \centering
  \newcommand{\logPlotScale}{0.9}
  \begin{subfigure}{0.45\linewidth}
    \scalebox{\logPlotScale}{
      \input{Help_5_log_good.pgf}
    }
    \caption{\centering A compliant monitored client.}
    \label{fig:compliant-client}
  \end{subfigure}
  \hfill
  \begin{subfigure}{0.45\linewidth}
    \scalebox{\logPlotScale}{
      \input{Help_5_log_bad.pgf}
    }
    \caption{\centering A non-compliant monitored client.}    
    \label{fig:noncompliant-client}
  \end{subfigure}
  \caption{Evolution of the monitoring status for the \lab{Help} branch of $\stS[game]$ (\Cref{eg:example-intro}).}
  \label{fig:live-logging}
\end{figure}

\begin{example}
  \label{eg:monitor-plots}
  The plots depicted in \Cref{fig:live-logging} are generated from monitor logs:
  they show the executions of two different clients for the guessing game PST $\stS[game]$ (\Cref{eg:example-intro}). The two clients select the \lab{Help} choice with different patterns and frequencies.
  The client in \Cref{fig:compliant-client} complies with the PST: it selects \lab{Help} with a frequency that always remains within the confidence interval of the probability specified in $\stS[game]$.
  Instead, the client in \Cref{fig:noncompliant-client} diverges from the PST: it selects \lab{Help} too often, hence the estimated probability moves outside the confidence interval after a few iterations; when this happens, the monitor issues a warning. 
  \qed
\end{example}

\section{Evaluation}\label{s:evaluation}

We evaluate \pstmonitor by measuring the overheads induced by the monitors it synthesises.
As a benchmark, we consider a probabilistic fragment of the Simple Mail Transfer Protocol (SMTP) \cite{SMTP} (server-side) formalised as the session type $\stS[smtp]$ below:
{\small
\begin{align}
  \stS[smtp] =\ & \stSnd{M220}{\typeStr}{}.\stBraOp\left\{\begin{array}{l}
    \stRcv{Helo}{\typeStr}{}.\stSnd{M250}{\typeStr}{}.\stS[mail],\\
    \stRcv{Quit}{}{}.\stSnd{M221}{\typeStr}{}
  \end{array}\right\}\nonumber\\[2mm]
  \label{s-mail-1}\stS[mail] =\ & \stRec{X}.\big(\stBra{\stRcv{MailFrom}{\typeStr}{0.5}.\stSnd{M250}{\typeStr}{}.\stRec{Y}.\\
  &\ \big(\stBraOp\left\{\begin{array}{l}
    \stRcv{RcptTo}{\typeStr}{0.6}.\stSnd{M250}{\typeStr}{}.\stRecVar{Y},\\
    \stRcv{Data}{}{}.\stSnd{M354}{\typeStr}{}.\stRcv{Content}{\typeStr}{}.\stSnd{M250}{\typeStr}{}.\stRecVar{X},\\
    \stRcv{Quit}{}{}.\stSnd{M221}{\typeStr}{}
  \end{array}\right\}\big),\nonumber\\
  &\quad \stRcv{Quit}{}{}.\stSnd{M221}{\typeStr}{}}\big)\nonumber
\end{align}
}%
When a client establishes a connection, the server sends a welcome message (\lab{M220}), and waits for the client to identify itself (\lab{Helo}). 
Then, the client can recursively send emails by specifying the sender (\lab{MailFrom}) and recipient (\lab{RcptTo}) address(es), followed by the mail contents (\lab{Data}). 
The client can send multiple emails by repeating the loop on ``$\stRecVar{X}$'' starting from line \eqref{s-mail-1}. 
The purpose of the probability annotations in $\stS[smtp]$ and $\stS[mail]$ is to flag clients that appear to be sending spam, or using the server resources without a purpose.
Setting a confidence level of 95\%, such probability annotations result in a warning if a client \begin{enumerate*}[label=\emph{(\alph*)}]
  \item sends three or more emails in a single connection (\lab{MailFrom}), or
  \item includes six or more recipients (\lab{RcptTo}).
\end{enumerate*}  

%
The type $\stS[smtp]$ above and the synthesised monitors are not tied to any specific message transport protocol; we implement them using TCP/IP (as per \cite{SMTP}) by providing suitable connection managers to the synthesised monitor (as explained in \Cref{sec:workflow}).
To conduct our experiments, we implement a client that sends emails to an SMTP server and measures the response time. 
As a server, we use a default instance of Postfix\footnote{\url{http://www.postfix.org/}} (one of the most used SMTP servers \cite{SMTPSurvey}) configured to receive emails and discard them. 
To study the monitoring overheads, we compare:
\begin{itemize}[nosep,noitemsep,label=-]
  \item \emph{an unsafe setup}: the SMTP client and server interact directly; 
  \item \emph{a black-box monitored setup} where communication is mediated by a monitor instantiated separately,
    as in \Cref{fig:wrapper-setup}; and 
  \item \emph{a grey-box monitored setup} with the client (written in Scala+\lchannels) and monitor running on the same virtual machine, as in \Cref{fig:hybrid-setup}.
\end{itemize}

\begin{figure}
  \newcommand{\plotScale}{0.65}
  \hspace{2em}\includegraphics[scale=0.2]{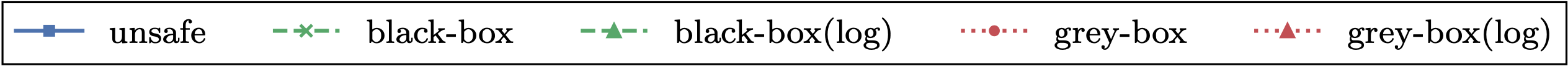}

  \smallskip
  
  \scalebox{\plotScale}{
    \input{smtp_cpu_consumption.pgf}
  }
  \scalebox{\plotScale}{
    \input{smtp_mem_consumption.pgf}
  }
  \scalebox{\plotScale}{
    \input{smtp_resp_time.pgf}
  }
  \caption{\centering Benchmark results for SMTP monitoring: CPU usage, maximum memory consumption, and response time. (Averages of 20 repetitions; 2 Xeon E5-2687W 8-core CPUs @ 3.10GHz, 128 GB RAM, GNU/Linux 5.10.27)}
  \label{fig:smtp-benchmarks}
\end{figure}

We measure the monitor's response time, CPU utilisation, and maximum memory consumption,
by running experiments where the client sends an increasing number of emails to the server.
We measure the performance with logging enabled and disabled (as explained in \Cref{s:monitoring-procedure}).
The results are shown in \Cref{fig:smtp-benchmarks}\footnote{
    Ideally, the results are also compared with other similar approaches. However, unfortunately, it is very hard to reliably compare overhead figures across different frameworks due to the discrepancies and peculiarities of each setup.}.

Overall, the plots in \Cref{fig:smtp-benchmarks} fall within the range of typical overheads introduced by run-time verification \cite{DBLP:series/lncs/BartocciFFR18, DBLP:journals/sttt/BartocciFBCDHJK19, DBLP:conf/fase/AcetoAFI21}.
They have two main origins:
\begin{enumerate}[nosep,label=\emph{(\arabic*)}]
  \item the calculation and bookkeeping of probability estimates at choice points, and their checking against $\stS[smtp]$; and
  \item the translation and duplication of messages being forwarded between the client and server (which is mitigated in the grey-box setup).
\end{enumerate}
The response time is arguably the most important measurement, since slower response times can be immediately perceived when interacting with a monitored system:  
in the black-box monitored setup we measured an overhead of 25\% (or 30\% with logging enabled).
Such overhead can be reduced to 13\% (or 20\% with logging enabled) by adopting a grey-box setup which minimises network communication.

\section{Conclusions and Discussion}\label{sec:conc}

We have presented \pstmonitor, a tool aimed to analyse quantitative aspects of a system's interactions at runtime. 

\pstmonitor is implemented as an extension of the monitoring framework \stmonitor from \cite{DBLP:conf/ecoop/BurloFS21,DBLP:journals/darts/BurloFS21}.
Originally, \stmonitor synthesises monitors from binary session types augmented with assertions (\ie predicates over communicated values). 
In order to support probabilistic session types (PSTs), we have replaced the type assertions with probabilities, and we have refactored the synthesis to generate monitors that conduct probabilistic analysis with the methodology introduced in \cite{DBLP:conf/coordination/BurloFSTT21} (and here summarised in \Cref{sec:workflow}).
We have also implemented the logging functionality outlined in \Cref{s:monitoring-procedure}, thus enabling further insight into the behaviour of a monitored system (as shown in \Cref{eg:monitor-plots}). 

Notably, the proposed methodology and the tool instantiating it are independent of each other. 
While the methodology can serve as the basis for other runtime analysis techniques for quantitative-based specifications, \pstmonitor is not limited, nor bound, to the current statistical technique. 
Rather, the synthesis permits for interchangeable statistical inference paradigms.
For instance, we can improve our CI estimation by utilising
the Wilson score interval \cite{Wilson1927} which is more costly but also more reliable when the sample size (observed messages up to the current
point of execution) is small or the specified probability is close to 0 or 1.

\paragraph{Related work}

Other publications and tools (also discussed in \cite{DBLP:conf/coordination/BurloFSTT21})
propose related techniques or have objectives similar to our work.
Albeit addressing the problem from a different angle with different specification languages, these tools and their respective methodologies can complement the analysis conducted by \pstmonitor. 

The work closest to ours is RT-MaC \cite{DBLP:conf/rtcsa/SammapunLS05}, a tool that generates monitors to determine whether a system satisfies a probabilistic property by analysing its behaviour at runtime and performing statistical hypothesis testing. 
Similarly to us, they make use of confidence intervals to gauge the accuracy of the observed behaviour. 
Unlike us, their tool sets up hypotheses from the specified (logic-based) properties and once the monitor has observed enough information to accept or reject the hypotheses, it decides whether the property is satisfied or not. 
By adopting the technique from \cite{DBLP:conf/rtcsa/SammapunLS05}, our monitors can be made to reach \emph{irrevocable verdicts} on the observed behaviour, rather than issue retractable warnings. 

Confidence intervals are also used in \cite{calinescu2015formal} for analysing quality of service properties (such as reliability, performance or cost) of a system. 
Unlike our work, their tool-supported approach is applied in the post-deployment phase on models obtained from system logs or via runtime monitoring. 
This is done with the aim of establishing unknown behavioural aspects of the system (\eg how often information is requested from a cache), and since such properties cannot be definitively confirmed with the available information (\ie a set of observations), confidence intervals are used to give an approximation of the system's behaviour. 
Similarly, LogLens \cite{debnath2018loglens} analyses system logs to detect anomalies; 
with no (or minimal) knowledge about the system, LogLens uses machine learning based techniques to discover patterns in its behaviour from previous system logs and then compare those in real-time logs to find inconsistencies. 
On the contrary, our technique does not rely on \emph{any} previous information about the system; moreover, the focus of \cite{calinescu2015formal,debnath2018loglens} is not on the probabilistic properties of the system. 
These tools can complement the methodology enabled by \pstmonitor: 
given their capability of logging information on the system executions (\Cref{s:monitoring-procedure}), our monitors can be used to extract information about the system itself (even when it is a black-box) which can be passed on to tools such as \cite{calinescu2015formal,debnath2018loglens} for further analysis. 

\bibliographystyle{elsarticle-num} 
\bibliography{refs}

\newpage
\section*{Current code version}
\label{}

\begin{table}[!h]
\begin{tabular}{|l|p{6.5cm}|p{6.5cm}|}
\hline
\textbf{Nr.} & \textbf{Code metadata description} & \textbf{Please fill in this column} \\
\hline
C1 & Current code version & v0.0.1 \\
\hline
C2 & Permanent link to code/repository used for this code version & \url{https://github.com/chrisbartoloburlo/stmonitor/tree/pstmonitor} \\
\hline
C3  & Permanent link to Reproducible Capsule & \\
\hline
C4 & Legal Code License   & BSD-2-Clause License \\
\hline
C5 & Code versioning system used & git \\
\hline
C6 & Software code languages, tools, and services used & Scala, Python \\
\hline
C7 & Compilation requirements, operating environments \& dependencies & \texttt{sbt} \\
\hline
C8 & If available Link to developer documentation/manual & \url{https://github.com/chrisbartoloburlo/stmonitor/blob/pstmonitor/README.md} \\
\hline
C9 & Support email for questions & \url{christian.bartolo@gssi.it} \\
\hline
\end{tabular}
\caption{Code metadata (mandatory)}
\label{} 
\end{table}

\listoffixmes

\end{document}